\documentclass[aps,prb,twocolumn,floatfix,amsmath,amssymb]{revtex4}
\usepackage[final]{graphicx}
\usepackage{bm}
\usepackage{float}
\usepackage{afterpage}

\begin{document}

\title{Interaction-driven phases in a Dirac Semimetal: Exact Diagonalization Results}

\author{Huaiming Guo$^{*}$ and Yongfei Jia}

\affiliation{Department of Physics, Beihang University, Beijing, 100191, China}

\begin{abstract}
The interaction-driven phases in the Dirac semimetal (SM) of the $\pi-$ flux model on square lattice are studied with nearest-(NN), next-nearest- (NNN) and next-next-nearest-neighbor (NNNN) interactions using the exact diagonalization method. We find that the NN interaction drives a phase transition from the SM phase to a charge density wave insulator. In the presence of the NNN interaction, the system becomes an anisotropic SM for small interactions and an insulator with the stripe order for large ones. The NNNN interaction drives the Dirac SM to a dimmerized insulator. The interplay of the NNN and NNNN interactions is also studied. It is found that the NNNN interaction firstly eliminates the effect of the NNN interaction and then develops its favorable order. In the calculations, the signature of the interaction-driven quantum anomalous Hall phase is not found.
\end{abstract}


\maketitle

\section{Introduction}
The discovery of topological insulators (TIs) has generated great interests in the field of condensed matter physics due to its many exotic electronic properties and many application potentials \cite{rev1,rev2,rev3,rev4}. Many efforts are devoted to the studies of TIs, among which the interplay between the interactions and the topological property is an important one \cite{int1}. The effects of the interactions on TIs have been extensively studied on different models. The consistent results are obtained using various analytical and numerical methods \cite{int1,kmu1,kmu2,th1,th2}.

Another important related problem is the possibility of the interaction-driven topological phase, which is firstly suggested on honeycomb lattice within the mean-field approximation \cite{tmi1}. It provides new approach to generate the topological phase without strong intrinsic spin-orbit coupling and will greatly extend the class of the topologically nontrivial materials. Though the phase is predicted on other models \cite{tmi2,tmi3,tmi4,tmi5}, the mechanism is still within the mean-field framework. So it is warranted to verify its existence with the exact methods. Recently there appears works addressing the problem using numerical exact diagonalization (ED), but inconsistent conclusions are made \cite{ed1,ed2,ed3,ed4}. Before large-scale numerical studies come out to clarify the problem, it is important to unify the results from the ED calculations.

Though the ED method is limited by its small size, it is an important method in dealing with the interacting systems. In the paper, based on the $\pi-$ flux model on square lattice, the interaction-driven phases in the Dirac semimetal (SM) are studied. The ED method with the momentum state as the basis is used, from which the momentum of the eigenstate is obtained. We show that it is very helpful in identifying the different quantum phases.  We systematically study the phases driven by the nearest-(NN), next-nearest- (NNN) and next-next-nearest-neighbor (NNNN) interactions.  We find that the NN interaction drives a phase transition from the SM phase to a charge density wave (CDW) insulator. In the presence of the NNN interaction, the system becomes an anisotropic SM for small interactions and an insulator with the stripe order for large ones. The NNNN interaction drives the Dirac SM to a dimmerized insulator. The interplay of the NNN and NNNN interactions is also studied. It is found that the NNNN interaction firstly eliminates the effect of the NNN interaction and then develops its favorable order. In the calculations, the signature of the interaction-driven quantum anomalous Hall (QAH) phase is not found.
\section{The model and method}
We consider a $\pi-$ flux model on square lattice with a tight-binding Hamiltonian \cite{ed1,model1},
\begin{equation}\label{eq1}
H_0= \sum_{ij}t_{ij}e^{i\chi_{ij}}c^\dag_{j}c_{i},
\end{equation}
where $c^\dag_{i}$ and $c_{i}$ are the annihilation and
creation operators at site ${\bf r}_i$.  For the case of  the sites $i$ and $j$ NN neighbors, $t_{ij}=t_1$ and $\chi_{i,i+\hat{x}}=0, \chi_{i,i+\hat{y}}=\pi i_x$. A unit cell contains two sites and in the reciprocal space, the Hamiltonian is written as $H_0=\sum_{\bf{k}}\psi_{\bf{k}}^{\dagger} {\cal H}_0(\bf{k})\psi_{\bf{k}}$ with $\psi_{\bf{k}}=(c_1,c_2)^{T}$ and ${\cal H}_0({\bf k})=2t_1\cos{k_x}\sigma_{x}-2t_1\cos{k_y}\sigma_{z}$, where $\sigma_{x,z}$  are the Pauli matrices. The energy spectrum is given by $E_{\bf k} = \pm \sqrt{4t_1^{2}(\cos^2 k_x+\cos^2 k_y)}$. The system is a SM with two inequivalent Dirac points at ${\bf K}_{1,2}=(\pi/2,\pm \pi/2)$. Before the effect of the interactions is studied, we firstly study the perturbations of the orders favored by the interaction to the Dirac SM.

The gapless Dirac points can be gapped by the NNN hopping with the pattern: $t_{ij}=t_2$ and $\chi_{i,i+\hat{x}+\hat{y}}=\chi_{i+\hat{x},i+\hat{y}}=\pi i_x$. In the momentum space it is: $ {\cal H}_{NNN}({\bf k})=-4t_2\sin k_x sin k_y \sigma_{y}$ with the energy spectrum
\begin{eqnarray*}
  E_{\bf k}^{(1)} = \pm \sqrt{4t_1^{2}(\cos^2 k_x+\cos^2 k_y)+16t_2^2\sin^2 k_x \sin^2 k_y}.
\end{eqnarray*}
 where ${\bf k}$ is in the reduced Brillouin zone $\{{\bf k}: |k_x|\leq \pi/2,|k_y|\leq \pi \}$. A nontrivial gap $4|t_2|$ opens at ${\bf K}_{1,2}$. The system is topological with gapless states associated with the edges traversing the gap and can be characterized by a nonzero Chern number.

The gapless Dirac points can also be gapped by the staggered CDW order $H_{cdw} = V_{c}\sum_{i}(-1)^{i_x+i_y}c^\dag_{i}c_{i}$.
Then a unit cell contains four sites. The Hamiltonian writes as
\begin{eqnarray*}
{\cal H}'_0({\bf k})&=&2t_1\cos{k_x}\sigma_{x}\otimes I+2t_1\cos{k_y}\sigma_{y}\otimes \sigma_{y}+V_{c} \sigma_{z}\otimes I,
\end{eqnarray*}
with the energy spectrum
\begin{eqnarray*}
  E_{\bf k}^{(2)} = \pm \sqrt{4t_1^{2}(\cos^2 k_x+\cos^2 k_y)+V_{c}^2},
\end{eqnarray*}
where ${\bf k}$ is in the reduced Brillouin zone $\{{\bf k}: |k_x|,|k_y|\leq \pi/2 \}$.
A gap $|V_{c}|$ opens at ${\bf K}_{1,2}$ and the resulting system is a trivial insulator.

We are also interested in the stripe order $H_{stripe} = V_{s}\sum_{i}(-1)^{i_x (i_y)}c^\dag_{i}c_{i}$. In the four-site unit cell as in the case of the CDW order, it writes as ${\cal H}_{stripe}({\bf k})=V_{s} \sigma_z\otimes \sigma_z$. The energy spectrum becomes,
\begin{eqnarray*}
  E_{\bf k}^{(3)} = \pm \sqrt{4t_1^{2}\cos^2 k_x+(2t_1\cos k_y\pm V_{s})^2}.
\end{eqnarray*}
For $|V_{s}|\leq |2t_1|$, the SM phase remains, but the touching points are anisotropic and are moved to other momenta. The anisotropic Dirac points can be gapped by the above NNN hopping, but not by the CDW order any more.

In the following of the paper, we use ED to study the effects of the NN, NNN and NNNN interactions on Eq.(\ref{eq1}),
\begin{eqnarray}
H_{int1}&=&V_1 \sum_{\langle ij\rangle} c^{\dagger}_i c_i, \\
H_{int2}&=&V_2 \sum_{\langle\langle ij\rangle\rangle} c^{\dagger}_i c_i, \\
H_{int3}&=&V_3 \sum_{\langle\langle\langle ij\rangle\rangle\rangle} c^{\dagger}_i c_i,
\end{eqnarray}
with $V_1, V_2, V_3$ the strengths of the interactions.
Since the total Hamiltonian is translationally invariant, the momentum states can be constructed as the basis of the ED calculations and the eigenenergies in each momentum sector are calculated. The momentum-dependent eigenenergies provide more information in distinguishing the interaction-driven quantum phases. In the following we set $t_1=1$ as the energy scale and all ED calculations are performed on $4\times 4$ system. The momentum is labeled by one integer $Q=k_x+N_x k_y$ with $N_x$ the number of unit cell in $x-$ direction (to include the above NNN hopping in some calculations, the unit cell with two sites along the $x-$ direction is chosen).
\section{The non-interacting SM phase}
\begin{figure}[htbp]
\centering
\includegraphics[width=8.cm]{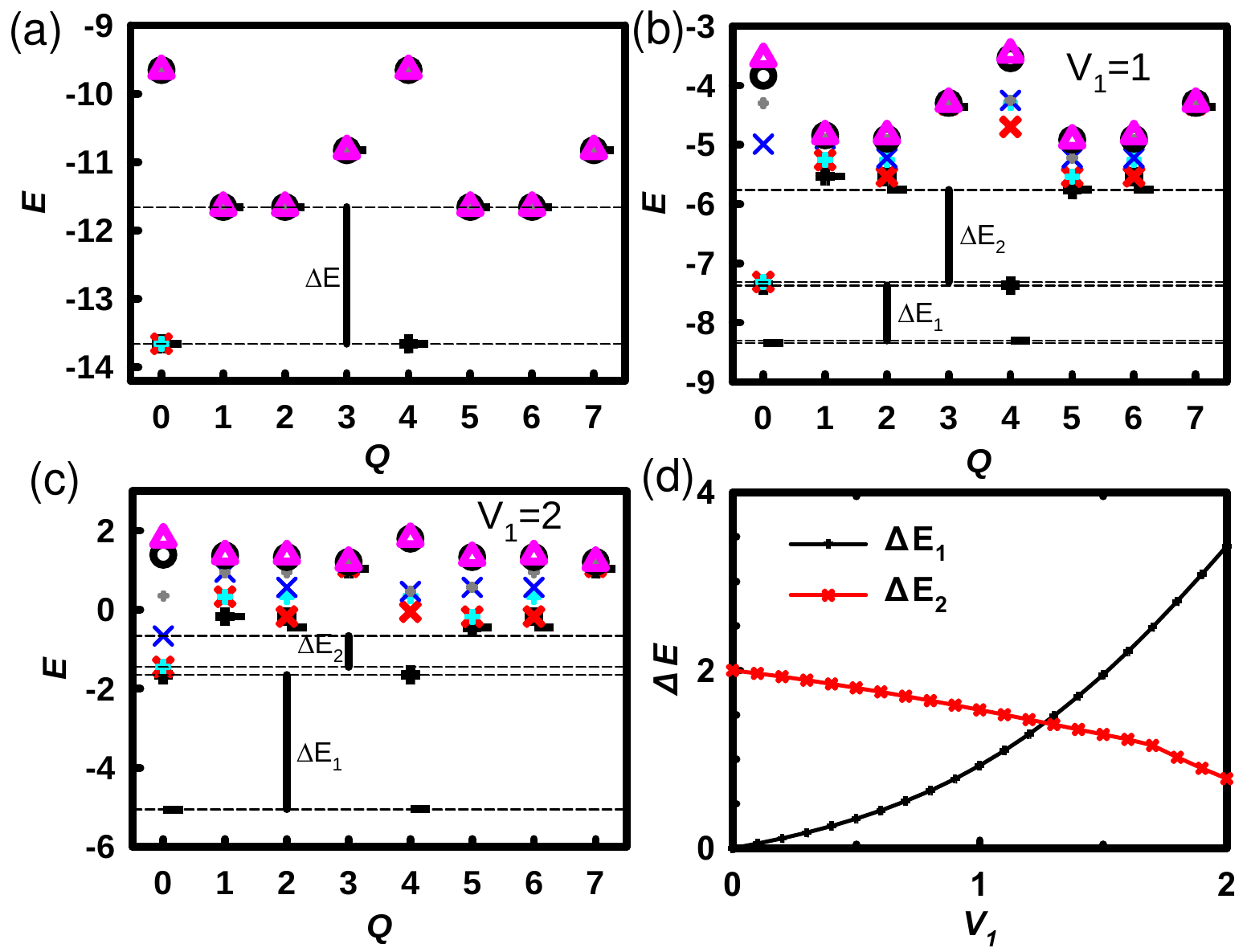}
\caption{(Color online) The energy spectrum in the momentum sector: (a) $V_1=0$; (b) $V_1=1$; (c) $V_1=2$. (d) $\Delta E_1$ and $\Delta E_2$ vs $V_1$.}\label{fig1}
\end{figure}

The main shortcoming of the ED calculation is the limited sizes. To understand the results on the small lattice, we firstly study the non-interacting SM phase using the ED method. The energy spectrum of the SM phase is shown in Fig.\ref{fig1}. At half filling the ground state is six-fold degenerate, four of which are in the momentum sector $(0,0)$ and two are in $(0,2)$. It is consistent with the band structure. For the $4\times 4$ lattice, the momenta are discrete as ${\bf k}=(k_x,k_y) \frac{\pi}{2}, k_x=0,1, k_y\in [0,3]$. Since the eigenenergy is two-fold degenerate at the Dirac points ${\bf K}_1=(1,1)$ and ${\bf K}_2=(1,2)$, two particles are chosen from the four states and the degeneracy is $6$. Out of the six cases, four of them have a particle at ${\bf K}_1$ and the other at ${\bf K}_2$, thus the four states have a total momentum $(0,0)$ or $Q=0$. Having both particles at ${\bf K}_1$ or ${\bf K}_2$ results in a total momentum $(0,2)$ or $Q=4$. Since the system is a SM, the energy spectrum is continuous. The gap between the six degenerate states and higher states ($\Delta E$ in Fig.\ref{fig1}) is due to the finite-size effect and will decrease as the sizes are larger.

If we use the above NNN hopping, CDW or stripe order to perturb the SM phase, the six-fold degenerate ground state is split and the ground state becomes non-degenerate. However for the case of the stripe order, it is known that the resulting phase is still SM, so the splitting is due to finite-size effect. While for the cases of the NNN hopping and the CDW order, the split non-degenerate ground state corresponds to a kind of insulating phase. So to identify the quantum phase in the system, it is important to distinguish the true gap and the finite-size gap. The problem can be solved by a finite-size scaling of the gap.

In the following calculations with the interactions, we also use the above orders with very small strength to probe the quantum phase of the ground state. Since the interacting Hamiltonian still has translational symmetry, the ground state obtained from the ED is multi-degenerate, which contains all the possible configurations of the quantum phase. The realistic ground state should be a spontaneous symmetry breaking one, which we generate by adding the above possible orders as a perturbation by hand. By this way the degenerate ground state is split slightly and the split one corresponds to a specific configuration.
\section{The effect of the NN interaction}
We firstly study the effect of the NN interaction described by Eq.(2) in the Dirac SM. The energy spectrums in the momentum sectors are shown in Fig.\ref{fig1}. When the NN interaction is added, the six-fold degenerate state is split into two groups, one of which contains two states and the other contains four states. There are two energy scales $\Delta E_1$ and $\Delta E_2$. For small interactions, it is expected that the ground state is still the Dirac SM \cite{robust}. So $\Delta E_1$ and $\Delta E_2$ are due to the finite-size effect. As the interaction is increased, $\Delta E_1$ is increased, while $\Delta E_2$ is decreased. So after a critical interaction, $\Delta E_1$ should become a true gap. Then the ground state is two-fold degenerate. The realistic ground state is a spontaneous symmetry breaking CDW insulator. So the NN interaction drives a phase transition from the SM phase to a CDW insulator.

There are two possible configurations for the CDW order, supposing they are: $\varphi_1, \varphi_2$. The momentum states are constructed from the representative states, using the translating operator ${\cal T}^{\bf r}$, which translates a state by the vector ${\bf r}$. Since some translating operators translate $\varphi_1$ ($\varphi_2$) to $\varphi_2$ ($\varphi_1$), one of $\varphi_1, \varphi_2$ is representative and we choose $\varphi_1$. The momentum state is \cite{ed},
\begin{eqnarray*}
\varphi'({\bf k})\propto \sum_{\bf r} e^{-i {\bf r}\cdot {\bf k}}{\cal T}^{\bf r} \varphi_{1}.
\end{eqnarray*}
It is straightforward that $\varphi'({\bf k})$ is nonzero at $(0,0)$ or $Q=0$ and $(0,2)$ or $Q=4$. The resulting momentum states are: $\varphi'_1=(\varphi_1+\varphi_2)/2$ and $\varphi'_2=(\varphi_1-\varphi_2)/2$, respectively. $\varphi'_{1 (2)}$ is translational invariant and is the state obtained in the ED calculations since the original interacting Hamiltonian has the translational symmetry. So the momentum of the state helps to distinguish them in the energy spectrum.

\section{The effect of the NNN interaction}

\begin{figure}[htbp]
\centering
\includegraphics[width=8.cm]{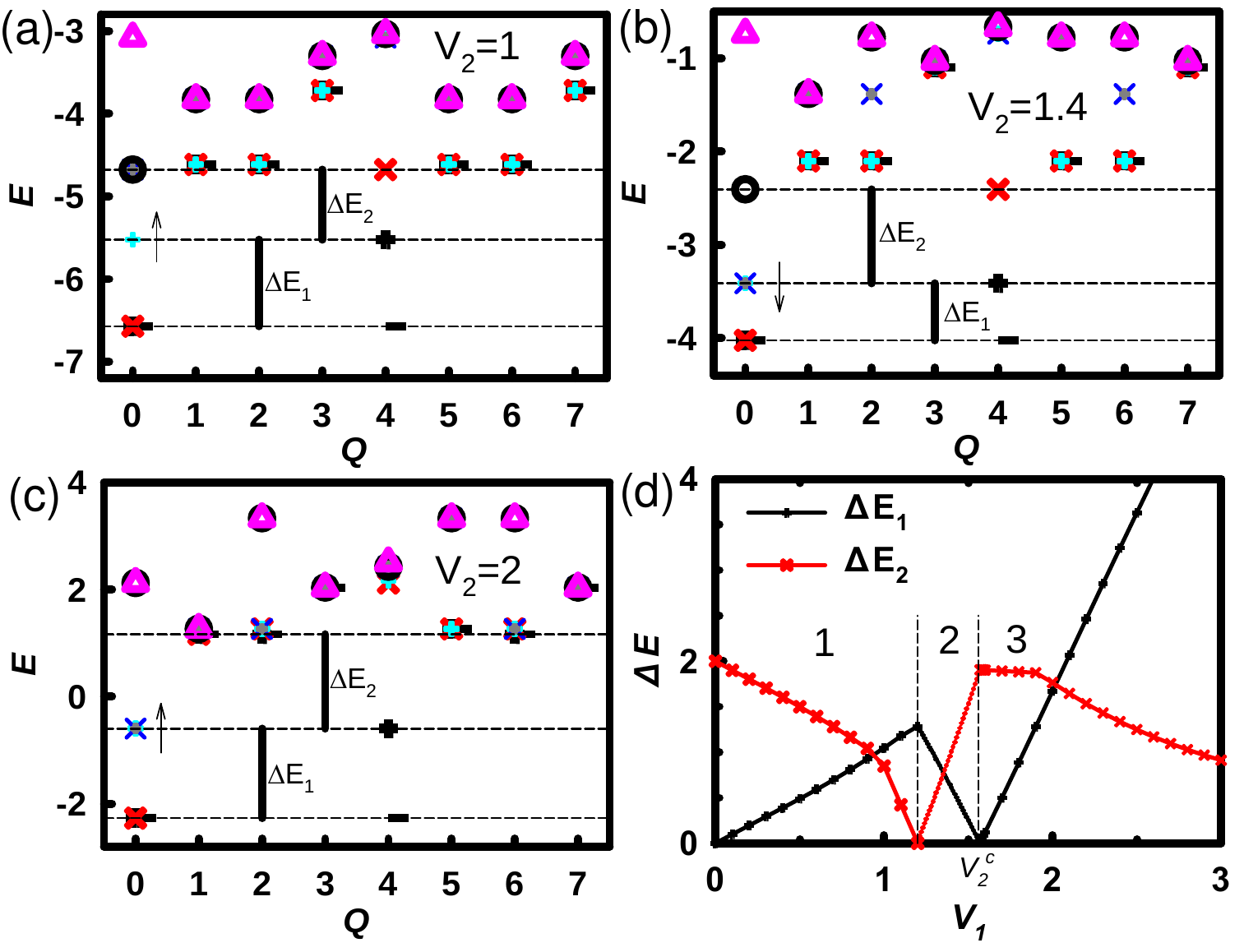}
\caption{(Color online) The energy spectrum in the momentum sector: (a) $V_2=1$; (b) $V_2=1.4$; (c) $V_2=2$. (c) $\Delta E_1$ and $\Delta E_2$ vs $V_2$. The arrows represent the moving direction of the nearby states as $V_2$ is increased. The three processes described in the text are denoted in (d).}\label{fig2}
\end{figure}

Next we study the effect of the NNN interaction described by Eq.(3) in the Dirac SM. The energy spectrums in the momentum sectors are shown in Fig.\ref{fig2}. As the interaction is small, the six-fold degenerate state is also split into two groups. The one containing four states has lower energy. There are two energy scales $\Delta E_1$ and $\Delta E_2$, too. Since the Dirac SM is robust to small interactions,  $\Delta E_1$ and $\Delta E_2$ are due to the finite-size effect for small interactions.

As $V_2$ is increased, there are the following processes in sequence (see Fig.\ref{fig2}): 1, $\Delta E_1$ increases and $\Delta E_2$ tends to vanish; 2, $\Delta E_1$ tends to vanish and $\Delta E_2$ increases (different to the process 1, two new states at $Q=0$ appears in the group with higher energy; 3, $\Delta E_1$ increases and $\Delta E_2$ tends to vanish.
In the process 3, since $\Delta E_1$ increases as $V_2$ increases, it is expected that at large $V_2$, $\Delta E_1$ is a true gap. The ground state is four-fold degenerate and has the stripe order, which can be probed by adding a very small specific stripe order to induce the spontaneous symmetry breaking. Also similar to the analysis in the previous section, three configurations of the stripe order are at $Q=0$ and one at $Q=4$.

So the phases driven by the small and large NNN interactions are identified. The more interesting region is the moderate NNN interactions, where an important question is whether the QAH topological phase is generated. Before addressing the question, we firstly study the properties of the low-energy states in the energy spectrum. Generally, the low-energy eigenstates can be classified into two groups. In the process 1, the upper group is two-fold degenerate. The two states can be probed by the CDW order. The lower group is four-fold degenerate, which can be probed by the stripe order or the above NNN hoppings. In the process 2 and 3, the upper and lower groups are all four-fold degenerate, both of which can be probed by the stripe order. In the process 2 (3), the lower (upper) group can also be probed by the NNN hoppings.

For the moderate NNN interaction, the system is in the process 2. The ground state is four-fold degenerate and can be probed by the stripe order or the above NNN hoppings, but not by the CDW order. It suggests that the system is the anisotropic Dirac SM, as the one described in Sec.2 when the static stripe order is added to a Dirac SM. So only the NNN interaction can not drive a Dirac SM to a QAH, which is consistent with the mean-field result.

\section{The effect of the NNNN interaction}
\begin{figure}[htbp]
\centering
\includegraphics[width=8.cm]{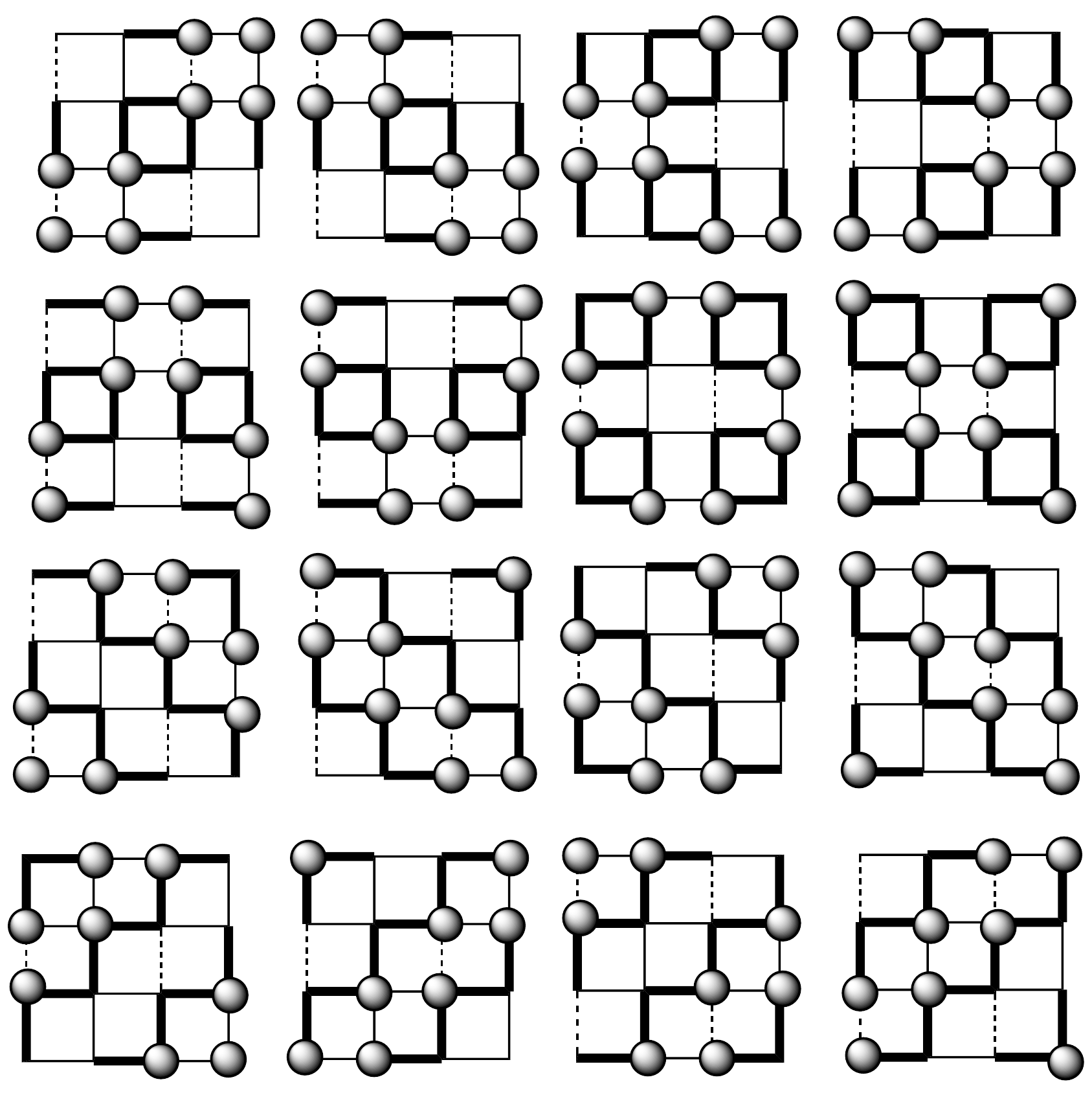}
\caption{The sixteen degenerate configurations favored by the NNNN interaction.}\label{fig3}
\end{figure}

In the mean-field approximation, the interaction-driven QAH phase needs to be stabilized by an additional NNNN interaction \cite{tmi4}. So in this section we study the effect of the NNNN interaction described by Eq.(4) and its interplay with the NNN interaction in the Dirac SM.

\begin{figure}[htbp]
\centering
\includegraphics[width=8.cm]{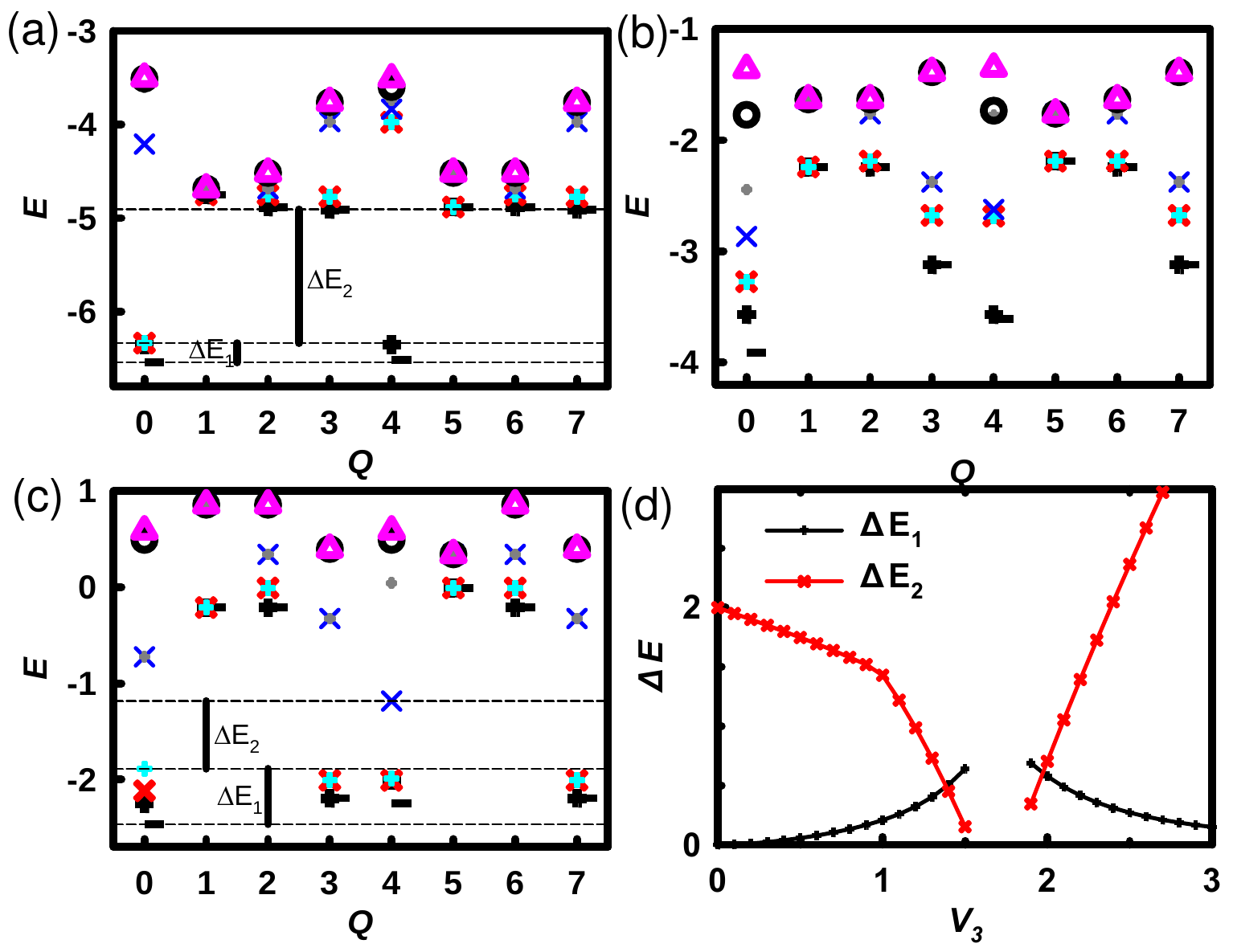}
\caption{(Color online) The energy spectrum in the momentum sector: (a) $V_3=1$; (b) $V_3=1.5$; (c) $V_3=2$. (d) $\Delta E_1$ and $\Delta E_2$ vs $V_3$. }\label{fig4}
\end{figure}

Firstly we study the effect of the sole NNNN interaction. In the atomic limit, the NNNN interaction stabilizes $16$ configurations (see Fig.\ref{fig3}), in which the interaction can be avoided. In the Dirac SM and for small interactions, the six-fold degenerate ground state is split with the energy difference $\Delta E_1$ (see Fig.\ref{fig4}). As the interaction is increased, $\Delta E_2$ decreases and $\Delta E_1$ increases. However since the Dirac SM is robust to small interactions, $\Delta E_1$ and $\Delta E_2$ are due to the finite-size effect. For moderate interactions, the energy scale is not obvious in the energy spectrum. As the interaction is further increased, sixteen states begin to evolve into the low-energy states. There are four states at $Q=0,3,4,7$, respectively. The momenta correspond to the momentum states of the configurations shown in Fig.\ref{fig3}. Here the energy scales $\Delta E_1$ and $\Delta E_2$ can be defined, with $\Delta E_1$ the splitting the sixteen low-energy states and $\Delta E_2$ the gap from the higher states. At large interaction and as it increases, $\Delta E_1$ decreases while $\Delta E_2$ increases, suggesting that the ground state is an insulator with the order shown in Fig.\ref{fig3}. Specially in the system with such kind of order, the NN hopping amplitudes form $8$ different patterns, in which the NN hopping amplitudes are dimmerized along the $x-$ and $y-$ directions. So the ED results suggest a phase transition from the Dirac SM to the insulator with the dimmerization driven by the NNNN interaction. However the detail of the phase transition is beyond the scope of the present ED method.

\begin{figure}[htbp]
\centering
\includegraphics[width=8.cm]{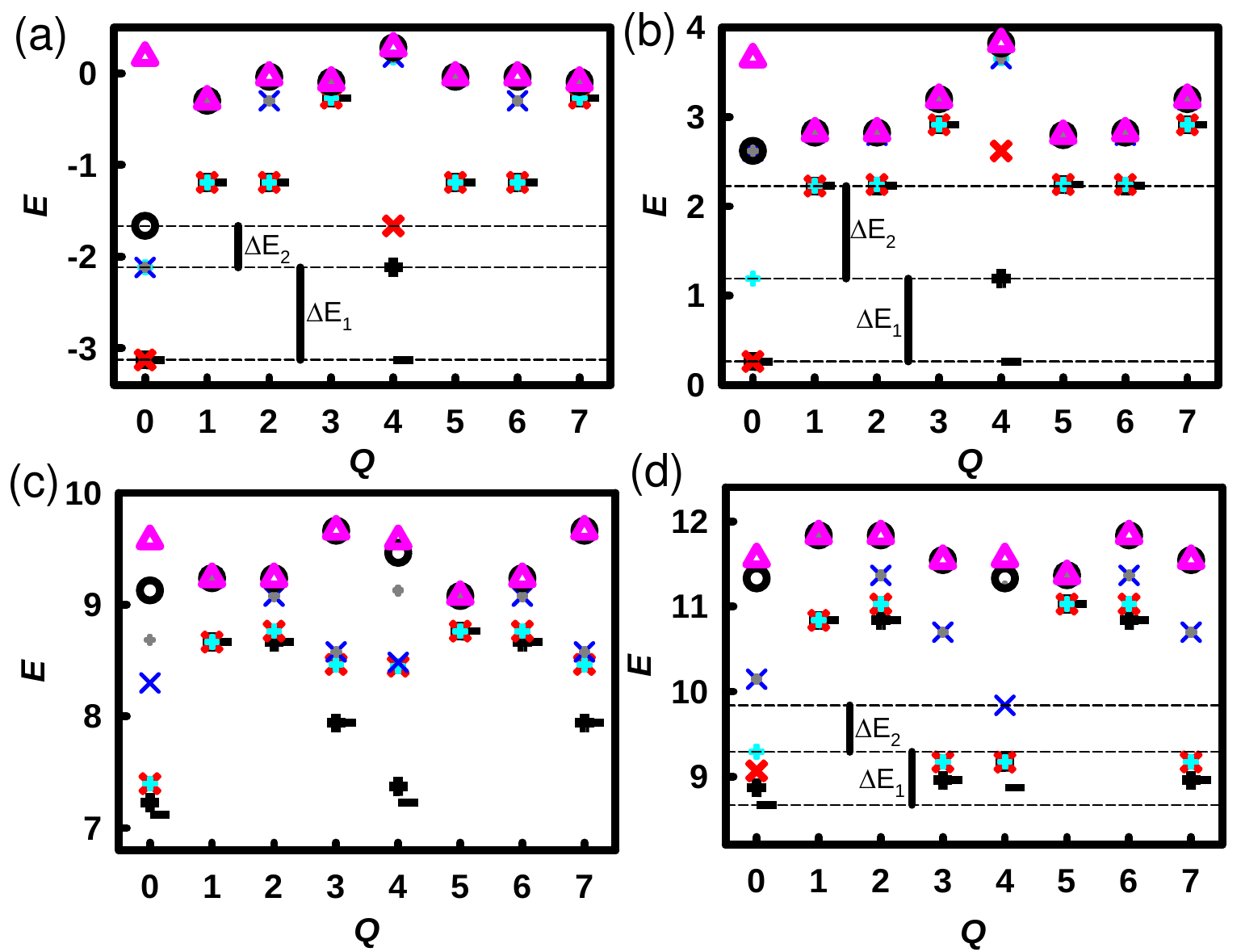}
\caption{(Color online) The energy spectrum in the momentum sector at fixed $V_2=1.4$: (a) $V_3=0.1$; (b) $V_3=0.5$; (c) $V_3=1.5$; (d) $V_3=2$. }\label{fig5}
\end{figure}

Next we study its interplay with the NNN interaction and the possible QAH phase driven by them. We add the NNNN interaction to a system with the NNN interaction. It is found that as the NNNN interaction is increased, it firstly eliminates the effect of the NNN interaction and then develop its favorable order. An example at $V_2=1.4$ is shown in Fig.\ref{fig5}.  As known in Fig.\ref{fig2} (b), when only the NNN interaction exists, the upper group containing four states begins to go down as the interaction is increased. After the NNNN interaction is added, it shows that the group begins to go up. Then the number of the states in the group becomes two and begins to go down. Finally at relatively large interactions, sixteen low-energy states are developed, which is favored by the NNNN interaction. We also perform the calculations at other values of $V_2$ and the results are the same.

It is suggested in the previous works that the QAH phase preserving the symmetry of the system should be two-fold degenerate. However in our calculations we find no such signature, which is consistent with our previous work \cite{ed1}.
\section{Conclusion and discussion}
We study the interaction-driven phases in the Dirac SM of the $\pi-$ flux model on square lattice. To properly identify the  quantum phases from the ED results, the effects of the static orders are firstly studied. We consider the staggered CDW, the stripe order and the nontrivial NNN hopping, which are favored by the considered interactions. Then we show that the non-interacting SM phase is characterized by a six-fold degenerate ground state, whose momenta are consistent with those from the analysis of the band structure. The gapping of the SM by the orders becomes the splitting of the degeneracy.

The effect of the NN interaction is firstly considered. We calculate the energy spectrum in the momentum sector. Though the results are affected by the finite-size effect, a phase transition from the SM phase to a CDW insulator is still identified. Next we study the effect of the NNN interaction. The results show that for small interactions, the SM phase is robust but becomes anisotropic; while for large interactions, it is an insulator with the stripe order.

To explore the interaction-driven QAH phase, we study the effect of the NNNN interaction. It is found that the sole NNNN interaction drives a phase transition from the SM to a dimmerized insulator. In the presence of the NNN interaction, its effect is found to be that it firstly eliminates the effect of the NNN interaction and then develops its favorable order. However the signature of the interaction-driven QAH is not found.

Finally we want to emphasize that the present results are in the scope of the ED method and are limited by the small sizes. Large-scale numerical calculations are warranted to verify them.
\section{Acknowledgements}
HG is supported by NSFC under Grant Nos. 11274032, 11104189, FOK YING TUNG EDUCATION FOUNDATION and Program for NCET.

\end{document}